\documentclass[conference]{IEEEconf}

\renewcommand\thesection{\arabic{section}} 
\renewcommand\thesubsectiondis{\thesection.\arabic{subsection}.}
\renewcommand\thesubsubsectiondis{\thesubsectiondis.\arabic{subsubsection}.}

\usepackage{amsmath,amssymb,amsfonts}
\usepackage{array}
\usepackage{url}
\usepackage{graphicx}
\def\BibTeX{{\rm B\kern-.05em{\sc i\kern-.025em b}\kern-.08em
    T\kern-.1667em\lower.7ex\hbox{E}\kern-.125emX}}
\usepackage{color, colortbl, soul}
\usepackage{colorspace}
\usepackage[utf8]{inputenc}

\usepackage[backend=biber,maxnames=6,minnames=1,style=ieee]{biblatex}
\AtEveryBibitem{%
  \iffieldundef{doi}{}{\clearfield{url}}
}

\newcommand{\records}{papers (post-merge)}
\newcommand{\Records}{Papers (post-merge)}

\makeatletter
\def\papers[#1]{\@nameuse{papers@#1}}
\def\source[#1]{\@nameuse{source@#1}}

\@namedef{source@acm}{4}
\@namedef{source@ieee}{50}
\@namedef{source@scopus}{277}
\@namedef{source@proceedings}{1}
\@namedef{source@other}{56}

\@namedef{papers@primary}{91}
\@namedef{papers@secondary}{20}
\@namedef{papers@excluded}{226}
\@namedef{papers@duplicate}{51}
\@namedef{papers@snowballing}{34}
\makeatother

\makeatletter
\@namedef{papers@excluded non-digital twins}{14}
\@namedef{papers@excluded missing diagnoses}{33}
\@namedef{papers@excluded cisportation}{81}
\@namedef{papers@excluded missing land route}{36}
\@namedef{papers@excluded no text}{21}
\@namedef{papers@excluded connection unclear}{10}
\@namedef{papers@excluded paywall}{19}
\@namedef{papers@excluded multiple times}{3}
\@namedef{papers@excluded paywall and other}{6}
\@namedef{papers@excluded secondary}{40}
\makeatother

\input{data/stats}
\input{data/stats2}
\newcommand{\bibsearchdate}{2022-07-15}

\usepackage[english]{datetime2}
\usepackage{enumitem}
\usepackage{longtable}
\usepackage{tabularx}

\usepackage{tikz}
\usepackage{pgfplots}
\usepackage{pgfplotstable}
\usepackage{pgfplotscolorcells}

\pgfplotsset{compat=1.17}
\usepgfplotslibrary{polar}
\usepackage{hyperref}
\usepackage[all]{hypcap}
\usepackage[noabbrev,capitalise]{cleveref}
\makeatletter
\@ifpackagewith{cleveref}{capitalise}{%
  \crefname{paragraph}{Paragraph}{Paragraphs}%
}{%
  \crefname{paragraph}{paragraph}{paragraphs}%
}
\makeatother

\newlist{questions}{enumerate}{1}
\setlist[questions]{label=RQ\arabic*:,ref=RQ\arabic*, leftmargin=!}
\newlist{incriteria}{enumerate}{1}
\setlist[incriteria]{label=IN\arabic*:,ref=IN\arabic*, leftmargin=!}
\newlist{excriteria}{enumerate}{1}
\setlist[excriteria]{label=EX\arabic*:,ref=EX\arabic*, leftmargin=!}
\setlist{labelindent=1.5em}

\DeclareCiteCommand{\acite}
{\usebibmacro{prenote}}
{%
  \usebibmacro{citeindex}%
  \printnames{labelname}%
  \setunit{\addspace}%
  \printtext[brackets]{%
    \bibhyperref{\printfield{labelnumber}}%
  }%
}
{\textcitedelim}
{\usebibmacro{postnote}}

\newcommand\copyrightnotice[1]{%
\begin{tikzpicture}[remember picture,overlay]
  \node[anchor=south,yshift=10pt] at (current page.south) {{\parbox{\dimexpr\textwidth-\fboxsep-\fboxrule\relax}%
      {\footnotesize #1}}};
\end{tikzpicture}%
}

\begin{document}
\title{A Systematic Mapping Study of Digital Twins for Diagnosis in Transportation}

\author{%
  Liliana M. Prikler$^{1,*}$ and Franz Wotawa$^{1}$\\
  \normalsize $^{1}$Graz University of Technology, Graz, Austria\\
  liliana.prikler@ist.tugraz.at, wotawa@ist.tugraz.at
  \normalsize *corresponding author
}

\maketitle

\begin{abstract}
  In recent years, digital twins have been proposed and implemented in
  various fields with potential applications ranging from prototyping to
  maintenance. Going forward, they are to enable numerous efficient
  and sustainable technologies, among them autonomous cars. However,
  despite a large body of research in many fields, academics have yet to
  agree on what exactly a digital twin is -- and as a result, what its
  capabilities and limitations might be. To further our understanding,
  we explore the capabilities of digital twins concerning diagnosis in
  the field of transportation. We conduct a systematic mapping study
  including digital twins of vehicles and their components, as well as
  transportation infrastructure.
  We discovered that few papers on digital twins describe any diagnostic
  process. Furthermore, most existing approaches appear limited to system
  monitoring or fault detection.
  These findings suggest that we need more research for diagnostic
  reasoning utilizing digital twins.
\end{abstract}
\IEEEoverridecommandlockouts

\copyrightnotice{
  \copyright{2023 IEEE.}
  Personal use of this material is permitted.
  Permission from IEEE must be obtained for all other uses,
  in any current or future media, including reprinting/republishing
  this material for advertising or promotional purposes,
  creating new collective works, for resale or redistribution to servers
  or lists, or reuse of any copyrighted component of this work in other
  works.
  The published version of this paper is available under the
  DOI \href{https://doi.org/10.1109/DSA59317.2023.00058}%
  {10.1109/DSA59317.2023.00058}.
}

\section{Introduction}

Transportation -- moving people or goods from one place to another -- is an essential process of many human-engineered systems. For example, we construct machines such as bicycles, cars, or trains to move people or goods faster, whereas streets, roads, and rails provide clear paths and surfaces to move on\footnote{For the remainder of this paper, we only consider transportation via land routes using such vehicles.}. However, engineered systems may fail and potentially lead to harm and health risks for involved people. For example, when an incident such as a car crash occurs, it is all too plain to see that it did, and what happened at the point of contact is also often visible in the remains.
However, even if the events leading up to the car crash are known,
the question of whether someone could prevent it often sparks an
entirely different debate.

The situation becomes even more critical when those systems gain autonomy. In contrast to ordinary cars, where a driver is able (and supposed to) intervene in case of a fault, no driver can take control of a fully autonomous vehicle without a steering wheel.
Hence, the car itself needs to perform a fully automated diagnosis of
any failure that might occur during operation -- from detecting that a failure did indeed occur to finding its root cause and finally repairing the affected part or mitigating the failure.
For mitigating a failure, a minimum-risk maneuver might be
carried out\footnote{Stopping immediately in case of a failure can be
  dangerous, e.g., on the middle of a highway or after a curve in
  a tunnel.
  Hence, the car must be able to carry out a more sophisticated maneuver
  without inducing gratuitous risks.}.
In the case of repair, the vehicle might use spare parts or
other means of redundancy to restore proper system functionality
for some duration.

In order to reason about the world around us -- both diagnostically and more broadly -- we rely on models that abstract complex behavior. However, due to simplification, these models are necessarily wrong, and sometimes (due to oversimplification) even harmfully wrong. The inferences we draw from such models can however only be as accurate as the models themselves. Therefore, it is believed that in order to only draw correct inferences, we need a model that fully replicates our subject of interest. This replica is the digital twin, and its behavior should mimic that of the subject, no matter which experiment is performed (cf.~\cite{grieves2017digital}).

This notion of a digital twin, while capturing the term's natural meaning, is however not a useful one. In practice, when researchers or engineers are tasked to create a digital twin, they inevitably apply their own judgment of ``realistic enough''. This judgment, in turn, may or may not be guided by metrics, such as how frequently we update the digital twin with actual measurements, how detailed it is, etc. (cf. \cite{jones2020characterising, vanderhorn2021digital}). Thus, for any particular ``digital twin'', the questions of whether it is a truthful reflection of the physical world and, more importantly, whether it is useful for drawing inferences remain. To find out, one typically has to study its applications in context.

Given the importance of diagnosis for preventing harm in the case of autonomous systems and the need for models (or ``digital twins'') to carry out reasoning tasks, we are interested in obtaining the current state of the art in research in this intersection.
It is worth noting that there exists already a sizable body of secondary
literature on digital twins, including reviews (systematic or otherwise),
surveys and gap analyses.
These reviews cover
manufacturing~\cite{kritzinger2018digital},
maintenance~\cite{errandonea2020digital},
safety management~\cite{agnusdei2021digital},
civil engineering~\cite{gao2021digital, jiang2021digital,
  bado2022digital},
batteries and battery management systems~\cite{wu2020battery,
  panwar2021recent, wang2021application, vandana2021multi},
electric and autonomous vehicles~\cite{vandana2021multi,
  bhatti2021towards, niaz2021autonomous},
and transportation~\cite{ali2022digital, qian2022digital}
among others.
However, we did not find any work that sufficiently discusses the issue
of diagnosis among them.
Therefore, we conduct a systematic mapping study of digital twins performing diagnosis on transport systems. In the study, we consider applications in vehicles and their components and supporting infrastructure with a diagnosis more concretely aimed at monitoring these systems, detecting and localizing faults within them, or deriving potential mitigation or repair strategies. With respect to these definitions, we aim to answer the following research questions:
\begin{questions}
\item\label{rq1}
  How many published research papers are available that discuss digital twins as a means of performing diagnosis in transportation systems?
\item\label{rq2}
  What are the types of digital twins published so far?
\item\label{rq3}
  How can trends and corresponding challenges be characterized?
\end{questions}

We structure the rest of this paper as follows:
In \cref{sec:methodology}, we describe the methodology we used to conduct our mapping study. In \cref{sec:applications,sec:types,sec:trends} we set out to answer the raised research questions. We start by listing the applications of digital twins in \cref{sec:applications}, then categorize them into different types in \cref{sec:types}, and finally discuss trends and issues in
\cref{sec:trends}. In \cref{sec:conclusion} we conclude our findings.

\section{Methodology}\label{sec:methodology}

We conducted a systematic mapping study following the guidelines of
\acite{kitchenham2007guidelines} as outlined in the following
paragraphs.

\paragraph{Search strategy}
We searched ACM, IEEE, and Scopus for papers containing at least one of
the keywords in each of the categories ``digital twin'', ``diagnosis'',
and ``transportation'' within their titles, abstracts or keywords.
The individual keywords were
\begin{description}
\item[digital twins] ``digital twin'', ``digital twins''
\item[diagnosis] ``diagnosis'', ``diagnostics'',  ``monitoring'', ``repair'',
  any pair of (``fault'', ``error'', ``anomaly'') \(\times\)
  (``detection'', ``localization'', ``isolation'', ``mitigation'')
\item[transportation] ``automotive'', ``automobile'', ``bus'', ``car'',
  ``cars'', ``railway'', ``road'', ``traffic'', ``transit'',
  ``transport'', ``transportation'', ``vehicle'', ``vehicles''
\end{description}

\paragraph{Study selection}
Having received an initial set of candidates from the above searches,
we applied the inclusion criteria
\begin{incriteria}
\item\label{incl:diag} title, keywords, or abstract refer to the
  application of digital twins to diagnosis (or similarly fault detection,
  etc.),
\item full text available in English (translations allowed), and
\item published in a peer-reviewed venue since \DTMdate{2012-01-01}
  or slated for publication as of \DTMdate{\bibsearchdate},
\end{incriteria}
and exclusion criteria
\begin{excriteria}
\item paper already considered (i.e., \ a duplicate),
\item\label{excl:no-clear-connection}
  none of the abstract, introduction, or conclusion connects the digital twin,
  diagnosis and a use case in transportation together\footnote{We added
    this criterion, because some abstracts were vague enough to
    fit\ \ref{incl:diag}, but the paper itself was actually unrelated.
    It compensates for the over-eagerness of the inclusion criteria
    after the full text has been obtained.},
\item said transportation is not facilitated through a vehicle,
\item said vehicle does not travel via land routes, such as
  streets, roads, tracks, etc.
\end{excriteria}
to filter relevant papers.

We applied backward and forward snowballing as outlined by
\acite{wohlin2014guidelines} with a preference towards using
COCI~\cite{heibi2019coci} and CROCI~\cite{heibi2019crowdsourcing}
to quickly retrieve DOIs whenever possible.  We minimized the overhead
incurred by potential rollbacks by annotating our bibliography entries
with a list of entries citing them.  These citation edges would then
form a graph, wherein any included paper could be traced back to
a paper within the initial set.

\paragraph{Data extraction}

Having collected all the papers that we would consider relevant with
reference to inclusion and exclusion criteria, we extracted both
bibliographic metadata and actual information (listed in
\cref{tab:extracted-info}) from them.
In our extraction process, we would sometimes discover that two papers
were describing the identical twin, potentially at different stages of
implementation.
In order not to skew our results, we merged these papers into a single
record.

\begin{table}[ht]
  \caption{Information extracted per paper. If multiple papers
    describe the same twin, then their information is shared.}%
  \label{tab:extracted-info}
  \centering
  \begin{tabularx}{\linewidth}{l|X}
    \textbf{Key} & \textbf{Value} \\
    \hline
    subject & What is being digitally twinned \\
    model & What information is captured \\
    purpose & For which end the digital twin is created \\
    method & How a diagnosis is performed \\
    storage &  In which digital space the twin lives\\
    lifecycle & When in the lifecycle of the subject, the digital twin
                is being used \\
    step & Which diagnosis steps are performed \\
    status & How close (roughly) this digital twin is
             to being realized \\
    note & Additional notes
  \end{tabularx}
\end{table}

\paragraph{Data synthesis}
We performed narrative and qualitative data synthesis.
While some of our research questions are quantitative
(e.g., \ concerning the number of papers published) or partially
quantitative (e.g., \ trends and issues),
the data found in the studies are overwhelmingly qualitative.
As a result, we can report little more than the number of papers
using similar approaches or having made similar findings in that
regard.

\paragraph{Limitations}
While we adhere to the principles outlined by
\acite{kitchenham2007guidelines}, some choices were made at various
steps in the study process that could significantly alter the results
produced. We discuss these choices below.

Our search strategy only mentions ``digital twin'' and its plural as
synonyms to digital twins, when in fact, synonyms such as ``mirrored
spaces''~\cite{grieves2017digital} or
``as-is building information model''~\cite{jiang2021digital} do exist.
We choose to follow a descriptive rather than prescriptive approach
here; reporting what other researchers label as digital twin rather than
what some wish to be labeled as such.

Our selection criteria specifically mention published articles and might
thus be prone to publication bias, as well as biases introduced by the
search engines used.  We do somewhat address these biases by scanning
conference proceedings, but (w.r.t. \acite{kitchenham2007guidelines})
lack the means to assess them otherwise.

While we did not perform any explicit study assessment, we will note
the failure to provide any of the data in \cref{tab:extracted-info} as
a potential lack. This lack should, however, not influence the results
we report beyond debates about the proper denominator for our
findings.

\section{Diagnostic Applications of Digital Twins}%
\label{sec:applications}

Our search yielded \source[acm] results from ACM,
\source[ieee] results from IEEE,
and \source[scopus] results from Scopus.
We found \source[other] further candidates through snowballing --
of which we included \papers[snowballing] --
and \source[proceedings] candidate through the scanning of proceedings.

We discarded \papers[duplicate] duplicates and \papers[excluded] papers
that did not fit our selection criteria.
Remaining were \papers[primary] primary sources and \papers[secondary]
reviews.
Of the primary papers, one did not provide any useful information
w.r.t. \cref{tab:extracted-info} -- thus, we effectively only
reviewed \total[papers] papers (\total[records] post-merge).

\begin{figure}[ht]
  \centering
  \begin{tikzpicture}
    \begin{axis}[
      x tick label style={
        /pgf/number format/1000 sep=},
      ylabel=\Records,
      ymajorgrids,
      enlargelimits=0.12,
      ybar,
      bar width=2em,
      ]
      \addplot [fill=red] coordinates {
        (2017,\byyear[2017])
        (2018,\byyear[2018])
        (2019,\byyear[2019])
        (2020,\byyear[2020])
        (2021,\byyear[2021])
        (2022,\byyear[2022])
      };
    \end{axis}
  \end{tikzpicture}
  \caption{Number of papers (post-merge) by year.  We shortened the time
    axis to the span 2017-2022, as no paper published before 2017
    matched the inclusion criteria.}%
  \label{fig:by-year}
\end{figure}
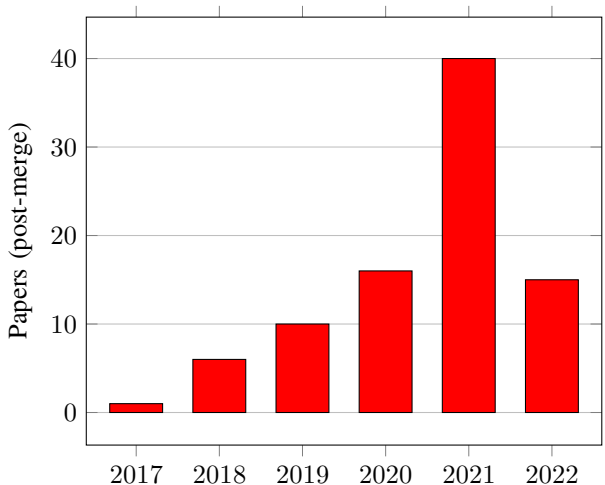

In \cref{fig:by-year} we give an overview of the number of papers by year.
Note, that we report numbers only for papers published since 2017
-- while we did indeed include papers published since 2012 in our search,
few results in the range from 2012 to 2016 were yielded, none of which
matched our selection criteria.  Further note, that we only provide the
post-merge count and that only papers within the same year have been
merged.

An overview of the subjects being digitally twinned is given in
\cref{tab:subject-summary}. Each subject falls within one or more
groups -- typically just one, but there are some instances in which the
boundaries are fluid, for example, the digital twin created in
\cite{errandonea2021iot}. Similarly, researchers may digitally twin
more than one subject for a given paper. For instance, they might
create several digital twins for vehicles and
pedestrians, which would result in a digital twin of a traffic
scenario.
In short, a single paper can fall into multiple groups
by either of the above.

\begin{table}[ht]
  \caption{Summary of digital twin subjects, also called ``real
    twins'' or ``physical twins'' in literature.}%
  \label{tab:subject-summary}
  \centering
  \begin{tabularx}{\linewidth}{p{8em}X}
    \textbf{Subject} & \textbf{Citations} \\
\hline
\multicolumn{2}{l}{\emph{components}} \\
\hline
batteries & \cite{baumann2018cloud} \cite{ramachandran2018recursive} \cite{merkle2019architecture} \cite{li2020digital} \cite{qu2020lithium} \cite{anandavel2021application} \cite{kortmann2021concept} \cite{merkle2021estimate} \cite{sancarlos2021from} \cite{singh2021implementation} \cite{olteanu2022battery} \cite{tang2022design} \cite{wang2022digital} \\
brakes & \cite{magargle2017simulation} \cite{rajesh2019digital} \cite{dygalo2020monitoring,dygalo2020real} \\
wheels & \cite{heckmann2020nonlinear} \cite{errandonea2021iot} \\
other & \cite{algin2018models} \cite{rassõlkin2019digital} \cite{venkatesan2019health} \cite{errandonea2021iot} \cite{kuts2021ros} \cite{rodríguez2021thermal} \cite{yang2021supercapacitor} \\
\hline
\multicolumn{2}{l}{\emph{environment}} \\
\hline
bridges & \cite{dang2018digital} \cite{forstner2019automated} \cite{shim2019digital} \cite{ye2019digital} \cite{bittencourt2021digital} \cite{dang2020bridge} \cite{dang2020master} \cite{sofia2020mobile} \cite{ye2020technology} \cite{bhouri2021model} \cite{dan2021digital} \cite{kang2021multimedia} \cite{yu2022digital} \cite{zhao2022developing} \\
railway & \cite{ahmadi2021adapting} \cite{avsievich2021railway} \cite{errandonea2021iot} \cite{kampczyk2021fundamental} \cite{sahal2021blockchain} \cite{zhang2021digital} \cite{bondoc2022employing} \cite{jiang2022research} \cite{zhou2022conceptual} \\
roads & \cite{machl2019planning} \cite{naets2019approach} \cite{li2021intelligent} \cite{meža2021digital} \cite{damico2022integrating} \cite{han2022bim} \\
traffic & \cite{radenkovic2020iot} \cite{tihanyi2021towards} \cite{wild2021joint} \cite{yun2021virtualization} \cite{hu2022digital} \cite{ma2022virtual} \cite{wang2022mobility} \cite{xu2022developing} \cite{yu2022digital} \\
other & \cite{zhang2019time} \cite{broekman2021real} \cite{clemen2021multi,lenfers2021improving} \cite{guo2021digital} \cite{zheng2021research} \\
\hline
\multicolumn{2}{l}{\emph{people}} \\
\hline
cyclists & \cite{clemen2021multi,lenfers2021improving} \cite{ma2022virtual} \\
drivers & \cite{kumar2018novel} \cite{liu2020sensor} \cite{wang2020digital} \\
human beings & \cite{wang2022mobility} \\
passengers & \cite{amrani2020architecture} \cite{campolo2020digital} \\
pedestrians & \cite{liu2020sensor} \cite{clemen2021multi,lenfers2021improving} \cite{guo2021digital} \cite{ma2022virtual} \\
\hline
\multicolumn{2}{l}{\emph{vehicles}} \\
\hline
automated guided vehicles & \cite{azangoo2021digital} \cite{braun2021robot} \\
cars & \cite{bai2021digital} \cite{yun2021virtualization} \cite{ma2022virtual} \\
haulers & \cite{javed2021safe} \\
trucks & \cite{shubenkova2018possibility} \\
trains & \cite{bustos2021methodology} \\
unspecified & \cite{kumar2018novel} \cite{naets2019approach} \cite{zhang2019time} \cite{amrani2020architecture} \cite{campolo2020digital} \cite{liu2020sensor} \cite{wang2020digital} \cite{azangoo2021digital} \cite{braun2021robot} \cite{guo2021digital} \cite{khaled2021streamlined} \cite{li2021intelligent} \cite{liu2021generative} \cite{marai2021roads} \cite{zhang2021energy} \cite{lv2022traffic} \cite{wang2022mobility} \cite{xu2022developing} \\
networks & \cite{zhao2020intelligent} \cite{liu2021distributed} \\

  \end{tabularx}
\end{table}

As we can see, most papers are interested in the ``environment'',
comprising infrastructure from the more commonly considered bridges,
railways and roads to the less frequent charging piles, bike stores and
subway stations, which fall under ``other'', as well as the
word ``infrastructure'' itself, ``traffic'' and the word ``environment''
itself, which also falls under ``other''.
Vehicle components (batteries, brakes and
wheels, as well as drive systems, motors, pantographs, powertrains,
propulsion drives and transmissions under ``other'')
are slightly less often digitally twinned than the
vehicles (cars, trains, trucks, as well as ``vehicle'')
themselves.\footnote{We also count digital twins of vehicular networks
  as digital twins of vehicles, as they are built on top of some model
  for the vehicles themselves. One could alternatively assign them
  towards infrastructure.}
The most petite group form people, which can be cyclists, drivers,
passengers, pedestrians, or simply ``human beings''.

Given this wide array of subjects, there is similarly a wide array of
purposes -- from estimating the state of charge or remaining useful life in
a battery to provide early safety warnings between a group of bridges.
For a proper analysis, we must therefore limit ourselves to each domain
in turn.

Starting with infrastructure, we find a number of applications for
bridges, as well as an early warning system for a bridge group
described in \cite{dan2021digital}.
While the wordings are indeed still very diverse, it becomes pretty clear
that interest is mostly focused on various aspects of structural health
monitoring, particularly under load.
Key purposes are detecting structural
damage~\cite{forstner2019automated}, assessing or
predicting the bridge's behaviour~\cite{dang2018digital,shim2019digital,
  dang2020master,ye2020technology,zhao2022developing}, predicting
its remaining life~\cite{yu2022digital}, and enabling new
maintenance paradigms such as preventive
maintenance~\cite{dang2020bridge}.
An interesting outlier is given in \cite{sofia2020mobile}.
While their work is performed on a bridge, they are concerned
about the flatness of the road on top.

\begin{table}[ht]
  \caption{Applications of digital twins in trains and railways.}%
  \label{tab:railway-applications}
  \centering
  \begin{tabularx}{\linewidth}{p{11.5em}X}
    \textbf{subject} & \textbf{purpose} \\
    \hline
    railway power system~\mbox{\cite{ahmadi2021adapting}} & decrease failure risk, predict destruction of electric equipment \\
railway tracks~\mbox{\cite{avsievich2021railway}} & determine track load, calculate structural stress \\
high-speed train~\mbox{\cite{bustos2021methodology}} & enable new maintenance paradigms \\
railway wheelset, railway pantograph~\mbox{\cite{errandonea2021iot}} & monitor state of railway infrastructure and vehicle, enable prognostic health management \\
railway turnouts~\mbox{\cite{kampczyk2021fundamental}} & estimate stress change \\
railway~\mbox{\cite{sahal2021blockchain}} & predictive maintenance \\
railway point machine~\mbox{\cite{zhang2021digital}} & enable safe train operations \\
light rail transit~\mbox{\cite{bondoc2022employing}} & forecast remaining useful life, prescribe maintenance plans \\
high-speed railway~\mbox{\cite{jiang2022research}} & improve infrastructure management, ensure safety \\
railway infrastructure~\mbox{\cite{zhou2022conceptual}} & integrate models into holistic platform

  \end{tabularx}
\end{table}

On the topic of roads, structural health monitoring reappears as
a purpose~\cite{meža2021digital}, combined with
detecting pavement distress~\cite{damico2022integrating}.
Finally, digital twins can also be applied
in the planning stage~\cite{machl2019planning} to determine
road use and load, and during road construction, for instance, to monitor compaction quality.~\cite{han2022bim}.

\Cref{tab:railway-applications} lists applications of digital twins
towards trains and railway infrastructure.
Estimating stress in tracks~\cite{avsievich2021railway} and
turnouts~\cite{kampczyk2021fundamental} appear as focuses on their own,
as does estimating the position and adhesion of
wheels~\cite{heckmann2020nonlinear}.
Researchers are also concerned about decreasing the risk of failure
and operating trains safely~\cite{ahmadi2021adapting,
  zhang2021digital}. Further, digital twins are to serve new
maintenance paradigms, such as predictive
maintenance~\cite{bustos2021methodology,sahal2021blockchain}.\footnote{
  \acite{errandonea2020digital} discuss the difference between predictive
  and preventive maintenance in greater detail.}
Within the context of railways, we also see that vehicle
and infrastructure is sometimes hard to separate.
As pointed out in \cite{errandonea2021iot}, detected faults can have
their origins in either the twinned component of the infrastructure
it interacts with, and more work is needed to figure out which of the
two is faulty.

In the domain of vehicle components, we find that digital twins of
batteries have very well defined (and often abbreviated)
common purposes.
Particularly, those are estimation of state of charge~\cite{
  ramachandran2018recursive,li2020digital,merkle2021estimate,
  sancarlos2021from,tang2022design,wang2022digital,
}, state of health~\cite{li2020digital,merkle2021estimate,
  wang2022digital},
state of power~\cite{wang2022digital}
remaining useful life~\cite{baumann2018cloud,anandavel2021application},
voltage~\cite{sancarlos2021from},
or parameter and state estimation in general~\cite{singh2021implementation}.
Battery digital twins are also used to evaluate~\cite{qu2020lithium}
and decelerate~\cite{anandavel2021application} performance degradation
as well as
``fulfill[ing] services for different stakeholders''~\cite{merkle2019architecture}.

Concerning other components, we again find digital twins used for estimating remaining useful life and
maintenance~\cite{magargle2017simulation,algin2018models,
  rajesh2019digital}.
As with batteries, estimation of the current state and parameters
is performed on powertrains~\cite{rodríguez2021thermal} and
supercapacitors~\cite{yang2021supercapacitor}.
In addition, however, we also see an application of digital twins
towards diagnosing the internal faults
of an automated brake system~\cite{dygalo2020monitoring,dygalo2020real}.

Once humans are introduced as an entity to be twinned, the focus
appears to shift towards predicting intentions.
In the context of transportation, this often concerns intentions and
behaviours of drivers~\cite{kumar2018novel,liu2020sensor,
  wang2022mobility}.
However, digital twins have also been proposed towards predicting
passenger flow in public transport~\cite{amrani2020architecture},
visualizing traffic~\cite{guo2021digital},
identifying and avoiding potential crashes~\cite{liu2020sensor,
  ma2022virtual}
and ``enabling autonomous vehicles''~\cite{marai2021roads,
  yun2021virtualization}.

Some of the applications of human digital twins are found in the
context of ``smart'' cities. Naturally, stakeholders such as
public transport agencies~\cite{campolo2020digital} and
bike rental services~\cite{clemen2021multi,lenfers2021improving} are
interested in reliable and valuable data for improving their services.
People also need to be surveilled as potential criminals or to
aid insurance companies~\cite{marai2021roads}. Sometimes, we may argue that this data is collected in the service of the public
interest. However, whether public interests are served and
whether this data collection can be justified needs to be
critically evaluated -- for a lengthier
discussion, see
\cref{sec:trends}.

\begin{table}[ht]
  \caption{Applications of vehicle digital twins (sans trains).}%
  \label{tab:vehicle-applications}
  \centering
  \begin{tabularx}{\linewidth}{p{9em}X}
    \textbf{subject} & \textbf{purpose} \\
    \hline
    trucks~\mbox{\cite{shubenkova2018possibility}} & predict failures, plan maintenance \\
electric vehicle, charging piles~\mbox{\cite{zhang2019time}} & evaluate efficiency of charging policies \\
software-defined vehicular networks~\mbox{\cite{zhao2020intelligent}} & verify routing schemes \\
automated guided vehicle, factory floor~\mbox{\cite{azangoo2021digital}} & reduce collisions between multiple vehicles \\
solar car~\mbox{\cite{bai2021digital}} & estimate energy consumption \\
automated guided vehicle~\mbox{\cite{braun2021robot}} & ease development of robots, raise alerts when robot and simulation behaviour diverge \\
autonomous haulers~\mbox{\cite{javed2021safe}} & identify potential platooning threats \\
hybrid electric vehicle~\mbox{\cite{khaled2021streamlined}} & notify drivers about failures \\
vehicles, road~\mbox{\cite{li2021intelligent}} & detect diversions, detect red light running, detect illegal parking \\
vehicular edge networks~\mbox{\cite{liu2021distributed}} & detect suspicious objects and behaviour \\
vehicles~\mbox{\cite{liu2021generative}} & detect GNSS attacks \\
agricultural vehicle~\mbox{\cite{woodcock2021uncertainty}} & correct steering direction \\
electric vehicle~\mbox{\cite{zhang2021energy}} & predict energy consumption \\
vehicles~\mbox{\cite{lv2022traffic}} & detect unsafe traffic events \\
vehicles, traffic~\mbox{\cite{xu2022developing}} & propose eco-routing strategies, mitigate carbon emissions

  \end{tabularx}
\end{table}

\Cref{tab:vehicle-applications} lists applications of digital twins
towards vehicles, with trains (already mentioned in \cref{tab:railway-applications}) and human digital twins being excluded.
Predicting failures and planning
maintenance~\cite{shubenkova2018possibility}, as well as notifying
drivers of failures that have occurred~\cite{khaled2021streamlined} are
mentioned as potential use cases for digital twins.
In electric vehicles, there is also some overlap between vehicle
digital twins and battery digital twins, as the former are also used
to estimate battery consumption~\cite{zhang2021energy} or evaluate the efficiency of charging policies~\cite{zhang2019time}.
As with human digital twins, surveillance of both legal and illegal
behavior is a further use case for vehicle digital
twins~\cite{li2021intelligent}.
In autonomous vehicles, however, there are also efforts to create digital
twins of other vehicles to monitor as traffic
participants~\cite{liu2020sensor,wang2020digital,yun2021virtualization}.

With some autonomous functions relying on vehicular networks for the
transmission of reliable data, traffic in the physical space is, however
not the only traffic to consider. In the context of vehicular networks,
digital twins may be used to define efficient routing
schemes~\cite{zhao2020intelligent}, but there is also a need to watch
out for attacks~\cite{liu2021generative} and more generally evaluate the
trustworthiness of other nodes within the
network~\cite{liu2021distributed}.

Some digital twins failed to match any of the previously mentioned
categories by means of subject matching.  These concern an
``environment'' or ``traffic'' itself, and are used to
estimate traffic~\cite{broekman2021real},
estimating speed and volume of traffic~\cite{hu2022digital},
or detecting pedestrians~\cite{tihanyi2021towards, wild2021joint}
and vehicles~\cite{wild2021joint}.

\section{Types of Digital Twins}\label{sec:types}

While some classifications of digital twins exist, few of them make
sense when trying to apply them towards diagnosis.
The distinction between digital twin prototypes and instances as
proposed by \acite{grieves2017digital} for instance, has little meaning
in this context.
While the diagnosis may be performed on prototypes to discover design flaws,
most meaningful applications -- specifically those at runtime -- require
instances.
The distinction between digital model, shadow, and twin introduced in \cite{kritzinger2018digital} might for some be an improvement
as it divides digital twins from somehow lesser technologies.
For diagnosis, however, a digital shadow would already be sufficient.
Furthermore, a digital twin without proper diagnosis should be
considered harmful, as faults injected in one part of the twin -- e.g.
faulty sensor readings -- would have severe implications for the digital
twin itself, as well as other digital twins and systems relying on it.

Therefore, we must accept that a digital twin refers to a model or a
collection of models researchers have labeled as such.  One of
the easiest categorizations is to split the twin back
up into its constituent models.
A summary of the models we found is given in \cref{tab:model-summary}.

\begin{table}[ht]
  \caption{Summary of models serving as a basis for a digital twin.}%
  \label{tab:model-summary}
  \centering
  \begin{tabularx}{\linewidth}{lX}
    \textbf{Model} & \textbf{Citations} \\
\hline
\multicolumn{1}{l}{\emph{behavioural models}} \\
\hline
agent-based models & \cite{clemen2021multi,lenfers2021improving} \cite{ma2022virtual} \\
other & \cite{baumann2018cloud} \cite{kumar2018novel} \cite{zhang2019time} \cite{campolo2020digital} \cite{javed2020enforcing} \cite{liu2020sensor} \cite{wang2020digital} \cite{ahmadi2021adapting} \cite{javed2021safe} \cite{khaled2021streamlined} \cite{liu2021distributed} \cite{zhang2021digital} \cite{zheng2021research} \cite{wang2022mobility} \cite{xu2022developing} \\
\hline
\multicolumn{2}{l}{\emph{geometric models}} \\
\hline
building information models & \cite{dang2018digital} \cite{ye2019digital} \cite{dang2020bridge} \cite{dang2020master} \cite{meža2021digital} \cite{damico2022integrating} \cite{han2022bim} \cite{jiang2022research} \\
other & \cite{forstner2019automated} \cite{machl2019planning} \cite{bittencourt2021digital} \cite{campolo2020digital} \cite{javed2020enforcing} \cite{liu2020sensor} \cite{sofia2020mobile} \cite{azangoo2021digital} \cite{bustos2021methodology} \cite{congress2021digital} \cite{guo2021digital} \cite{kuts2021ros} \cite{li2021intelligent} \cite{liu2021generative} \cite{tihanyi2021towards} \cite{yun2021virtualization} \cite{zheng2021research} \cite{ma2022virtual} \\
\hline
\multicolumn{2}{l}{\emph{mathematical models}} \\
\hline
finite element models & \cite{magargle2017simulation} \cite{dang2018digital} \cite{shim2019digital} \cite{ye2019digital} \cite{dang2020bridge} \cite{ye2020technology} \cite{avsievich2021railway} \cite{bhouri2021model} \cite{bustos2021methodology} \cite{dan2021digital} \cite{bondoc2022employing} \cite{yu2022digital} \\
other & \cite{algin2018models} \cite{venkatesan2019health} \cite{dygalo2020monitoring,dygalo2020real} \cite{bai2021digital} \cite{cherepov2021methodology} \cite{kampczyk2021fundamental} \cite{lv2022traffic} \\
\hline
\multicolumn{2}{l}{\emph{statistical models}} \\
\hline
hidden Markov models & \cite{zhao2020intelligent} \\
other & \cite{algin2018models} \cite{ye2019digital} \cite{amrani2020architecture} \cite{liu2021distributed} \cite{yu2022digital} \\
\hline
\multicolumn{2}{l}{\emph{physics-based models}} \\
\hline
equivalent circuit models & \cite{baumann2018cloud} \cite{ramachandran2018recursive} \cite{li2020digital} \cite{kortmann2021concept} \cite{merkle2021estimate} \cite{singh2021implementation} \cite{yang2021supercapacitor} \cite{olteanu2022battery} \cite{tang2022design} \cite{wang2022digital} \\
Doyle-Fuller-Newman models & \cite{sancarlos2021from} \cite{singh2021implementation} \\
lumped-element thermal models & \cite{baumann2018cloud} \cite{rodríguez2021thermal} \\
mass-spring-damper models & \cite{naets2019approach} \\
other & \cite{magargle2017simulation} \cite{rajesh2019digital} \cite{zhang2019time} \cite{heckmann2020nonlinear} \cite{anandavel2021application} \cite{bhouri2021model} \cite{braun2021robot} \cite{kang2021multimedia} \cite{kortmann2021concept} \cite{woodcock2021uncertainty} \cite{zhang2021energy} \cite{zhao2022developing} \cite{zhou2022conceptual} \\
\hline
\multicolumn{2}{l}{\emph{other models}} \\
\hline
data models & \cite{shubenkova2018possibility} \cite{forstner2019automated} \cite{amrani2020architecture} \cite{bittencourt2021digital} \cite{booyse2020deep} \cite{qu2020lithium} \cite{radenkovic2020iot} \cite{sofia2020mobile} \cite{bai2021digital} \cite{broekman2021real} \cite{errandonea2021iot} \cite{kang2021multimedia} \cite{kuts2021ros} \cite{marai2021roads} \cite{sahal2021blockchain} \cite{sancarlos2021from} \cite{wild2021joint} \cite{zheng2021research} \cite{hu2022digital} \cite{jiang2022research} \cite{wang2022mobility} \cite{zhou2022conceptual} \\
structural models & \cite{algin2018models} \cite{azangoo2021digital} \cite{bustos2021methodology} \\

  \end{tabularx}
\end{table}

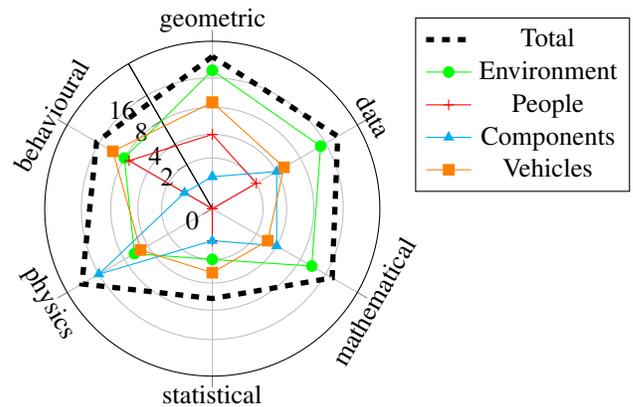
\begin{figure}[ht]
  \centering
  \begin{tikzpicture}
    \begin{polaraxis}[
      scale = 0.65,
      legend pos=outer north east,
      xmin=120,xmax=480,
      xtick={450,390,330,270,210,150},
      xticklabels={geometric,data,mathematical,%
        statistical,physics,behavioural},
      xticklabel style={sloped like x axis},
      ytick={0,2,4,8,16},
      yticklabels={0,2,4,8,16},
      y coord trafo/.code =\pgfmathparse{log2(1+#1)},
      y coord inv trafo/.code=\pgfmathparse{2^#1-1},
      ]
      \addplot [line width=2pt, dashed] coordinates {
        ( 90, \model[geometric aggregated])
        ( 150,\model[behavioural aggregated])
        ( 210,\model[physics aggregated])
        ( 270,\model[statistical aggregated])
        ( 330,\model[mathematical aggregated])
        ( 390,\model[data])
        ( 450,\model[geometric aggregated])
      };
      \addplot [
      green,mark=*
      ] coordinates {
        ( 90, \model[environment geometric aggregated])
        ( 150,\model[environment behavioural aggregated])
        ( 210,\model[environment physics aggregated])
        ( 270,\model[environment statistical aggregated])
        ( 330,\model[environment mathematical aggregated])
        ( 390,\model[environment data])
        ( 450,\model[environment geometric aggregated])
      };
      \addplot [
      red,mark=+
      ] coordinates {
        ( 90, \model[people geometric aggregated])
        ( 150,\model[people behavioural aggregated])
        ( 210,\model[people physics aggregated])
        ( 270,\model[people statistical aggregated])
        ( 330,\model[people mathematical aggregated])
        ( 390,\model[people data])
        ( 450,\model[people geometric aggregated])
      };
      \addplot [
      cyan,mark=triangle*
      ] coordinates {
        ( 90, \model[component geometric aggregated])
        ( 150,\model[component behavioural aggregated])
        ( 210,\model[component physics aggregated])
        ( 270,\model[component statistical aggregated])
        ( 330,\model[component mathematical aggregated])
        ( 390,\model[component data])
        ( 450,\model[component geometric aggregated])
      };
      \addplot [
      orange,mark=square*
      ] coordinates {
        ( 90, \model[vehicle geometric aggregated])
        ( 150,\model[vehicle behavioural aggregated])
        ( 210,\model[vehicle physics aggregated])
        ( 270,\model[vehicle statistical aggregated])
        ( 330,\model[vehicle mathematical aggregated])
        ( 390,\model[vehicle data])
        ( 450,\model[vehicle geometric aggregated])
      };
      \legend{Total,Environment,People,Components,Vehicles}
    \end{polaraxis}
  \end{tikzpicture}
  \caption{Models of digital twins split by application area.}%
  \label{fig:models-by-application}
\end{figure}

We can draw apparent correlations between the context in which the digital twin is created and the models used for the purposes it serves.
As seen in \cref{fig:models-by-application}, environmental models
are mostly geometric (including many building information models), but
mathematical models such as finite element models serve similar
purposes.  Physics models, including various models of
batteries, are the most pronounced in components.  Considering humans and vehicles,
geometry -- that is, where they are, how large they are, and where they
move to -- and behavior are the essential information.
All of the above may also be available in data models, which are often
used for machine learning or are themselves machine learning
models.\footnote{We distinguish between mathematical and physical
  and mathematical, statistical and data models based on seemingly
  arbitrary lines drawn from the model descriptions given by these
  papers.  To us, it makes a qualitative difference whether a machine
  learning algorithm is fed ``a bunch of data'' or a finite-element
  model.
}

As digital twins consist of one or more models, diagnostic algorithms
that can be applied to these individual models should be applicable
to the digital twin.
Though seemingly obvious, this conclusion need not necessarily hold,
however.  Constraints about real-time performance might make
some algorithms infeasible.  Algorithms that work well on small data
sets but have lousy performance with high amounts of data also break
down.
To counter this drawback, surrogate models learned from the digital
twin could be used in place of the actual twin~\cite{kang2021multimedia}.
However, even if those algorithms remained applicable without learning
surrogate models, it would not be an achievement of the digital twin.
To be useful, digital twins would have to power new applications in
diagnosis that were previously unheard of or thought
infeasible.\footnote{Failing that, they should, at the very least, improve
  the efficiency of existing methods to offset the additional
  costs of creating and maintaining the digital twin.}

Two key points are essential for these applications: the diagnostic
use and the diagnosis methods.  The diagnostic use roughly describes
the capabilities of a digital twin.  Is it used for monitoring?
Can it detect faults?  Can it tell us where those faults lie?
Do we have to resort to expert knowledge to fix those faults, or does
the digital twin offer solutions?\footnote{If the digital twin
  does provide solutions, we might further inquire how well we could
  reason about those, and so on.}  The diagnosis methods, on the other
hand, are the algorithms employed to make the digital twin application
arrive at a given conclusion.

\Cref{tab:step-by-x} shows the correlation of models and methods
with diagnostic use.  Regarding models, we see that mathematical and
behavioral models are more closely associated with fault detection
than mere monitoring.  Both of them also score high in fault
localization.  For fault mitigation, there is no clear ``winner''.
While geometric and data models are tied for first place, and
behavioral and statistical models for second respectively, only
statistical methods are well-represented concerning their total diagnostic use.

Regarding methods, machine learning peaks in
monitoring and fault detection, but due to it being the single most used
technology, it is also the mainly used mitigation method.  It serves no
purpose in the localization of faults, though, which is most
often done through simulations or visualizations.  Signal processing is
most often used for monitoring but might also have applications for fault
detection.  All other methods are minor in occurrence.

\begin{table*}[ht]
  \caption{Correlation of digital twin models and diagnosis methods with
    diagnostic use.
    Note that a single model or technique might be applied for
    multiple purposes, for example, both fault detection and
    localization.
    The caveats from \cref{fig:digital-twin-method} also apply.
  }%
  \label{tab:step-by-x}
  \scriptsize
  \begin{filecontents*}{data/model_step.dat}
r,monitoring,fault detection,fault localization,fault mitigation
artificial intelligence,18,13,0,3
machine learning,17,12,0,3
simulation or visualization,9,11,5,3
signal processing,12,1,0,0
fault trees,0,2,0,1
thresholds,0,4,1,0
other,5,6,1,3
unknown,7,2,1,0
\end{filecontents*}
\pgfplotstabletypeset[
  col sep=comma,
  every column/.style={column type=p{5em}},
  columns/r/.style={column type=p{8em},
    column name={\emph{use by model}},
    string type},
  columns/monitoring/.style={
    color cells={low=0,high=15}
  },
  columns/fault detection/.style={column name={\mbox{fault}
      \mbox{detection}},
    color cells={low=0,high=10}
    },
  columns/fault localization/.style={column name={\mbox{fault}
      \mbox{localization}},
      color cells={low=0,high=3}
    },
  columns/fault mitigation/.style={column name={\mbox{fault}
      \mbox{mitigation}},
    color cells={low=0,high=3}
  },
  /pgfplots/colormap={summer}{rgb255=(0,128,102) rgb255=(255,255,102)}%
  ]{data/model_step.dat}
  \begin{filecontents*}{data/method_step.dat}
r,monitoring,fault detection,fault localization,fault mitigation
artificial intelligence,18,13,0,3
machine learning,17,12,0,3
simulation or visualization,9,11,5,3
signal processing,12,1,0,0
fault trees,0,2,0,1
thresholds,0,4,1,0
other,5,6,1,3
unknown,7,2,1,0
\end{filecontents*}
\pgfplotstabletypeset[col sep=comma,
  every column/.style={column type=p{5em}},
  columns/r/.style={column type=p{8em},
    column name={\emph{use by method}},
    string type},
  columns/monitoring/.style={
    color cells={low=0,high=10}
  },
  columns/fault detection/.style={column name={\mbox{fault}
      \mbox{detection}},
    color cells={low=0,high=10}
    },
  columns/fault localization/.style={column name={\mbox{fault}
      \mbox{localization}},
      color cells={low=0,high=3}
    },
  columns/fault mitigation/.style={column name={\mbox{fault}
      \mbox{mitigation}},
    color cells={low=0,high=3}
  },
  /pgfplots/colormap={summer}{rgb255=(0,128,102) rgb255=(255,255,102)}%
  ]{data/method_step.dat}
\end{table*}

\begin{figure}[ht]
  \centering
  \begin{tikzpicture}
    \begin{polaraxis}[
      scale = 0.8, every node/.style={scale=0.8},
      font = \scriptsize,
      legend pos=outer north east,
      xmin=110,xmax=470,
      xtick={450,410,370,330,290,250,210,170,130},
      xticklabels={artificial \mbox{intelligence},%
        machine learning,%
        numerical analysis,%
        simulation and visualization,%
        signal processing,%
        fault trees,%
        thresholds,%
        other,%
        unknown,%
      },
      xticklabel style={sloped like x axis, align=center,
        text width=8em},
      ytick={0,2,4,8,16,32},
      yticklabels={0,2,4,8,16,32},
      y coord trafo/.code =\pgfmathparse{log2(1+#1)},
      y coord inv trafo/.code=\pgfmathparse{2^#1-1},
      ]
      \addplot [line width=2pt, dashed] coordinates {
        (450,\method[artificial intelligence])
        (410,\method[machine learning])
        (370,\method[numerical analysis])
        (330,\method[simulation and visualization])
        (290,\method[signal processing])
        (250,\method[fault trees])
        (210,\method[thresholds])
        (170,\method[other])
        (130,\method[unknown])
        (90, \method[artificial intelligence])
      };
      \addplot [
      green,mark=*
      ] coordinates {
        (450,\method[environment artificial intelligence])
        (410,\method[environment machine learning])
        (370,\method[environment numerical analysis])
        (330,\method[environment simulation and visualization])
        (290,\method[environment signal processing])
        (250,\method[environment fault trees])
        (210,\method[environment thresholds])
        (170,\method[environment other])
        (130,\method[environment unknown])
        (90, \method[environment artificial intelligence])
      };
      \addplot [
      red,mark=+
      ] coordinates {
        (450,\method[people artificial intelligence])
        (410,\method[people machine learning])
        (370,\method[people numerical analysis])
        (330,\method[people simulation and visualization])
        (290,\method[people signal processing])
        (250,\method[people fault trees])
        (210,\method[people thresholds])
        (170,\method[people other])
        (130,\method[people unknown])
        (90, \method[people artificial intelligence])
      };
      \addplot [
      cyan,mark=triangle*
      ] coordinates {
        (450,\method[component artificial intelligence])
        (410,\method[component machine learning])
        (370,\method[component numerical analysis])
        (330,\method[component simulation and visualization])
        (290,\method[component signal processing])
        (250,\method[component fault trees])
        (210,\method[component thresholds])
        (170,\method[component other])
        (130,\method[component unknown])
        (90, \method[component artificial intelligence])
      };
      \addplot [
      orange,mark=square*
      ] coordinates {
        (450,\method[vehicle artificial intelligence])
        (410,\method[vehicle machine learning])
        (370,\method[vehicle numerical analysis])
        (330,\method[vehicle simulation and visualization])
        (290,\method[vehicle signal processing])
        (250,\method[vehicle fault trees])
        (210,\method[vehicle thresholds])
        (170,\method[vehicle other])
        (130,\method[vehicle unknown])
        (90, \method[vehicle artificial intelligence])
      };
      \legend{Total,Environment,People,Components,Vehicles}
    \end{polaraxis}
  \end{tikzpicture}
  \caption{Methods used to perform diagnosis on digital twins.
    Note that ``artificial intelligence'' subsumes machine
    learning, but also contains other forms of artificial intelligence
    (for example, rule-based systems).
    Some machine learning models additionally count towards
    signal processing by their usage.}%
  \label{fig:digital-twin-method}
\end{figure}
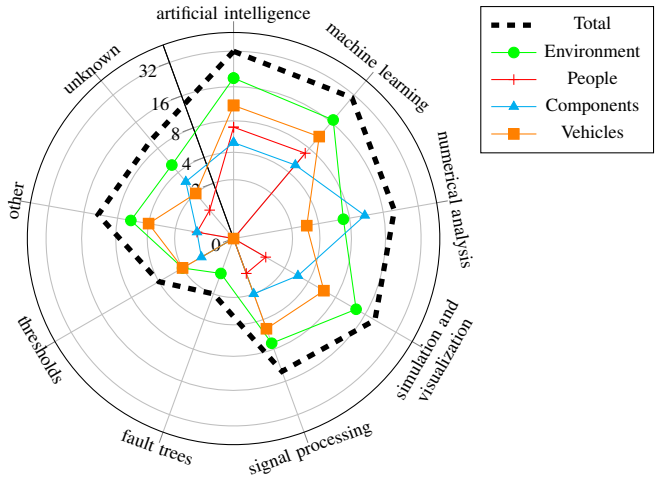

\Cref{fig:digital-twin-method} summarizes the methods used for
performing diagnosis on digital twins.  The most significant chunk is
unsurprisingly taken by artificial intelligence with a large focus on
machine learning systems.
Besides umbrella terms such as machine learning, deep learning, or
reinforcement learning, specific techniques such as support vector
machines, neural networks (including long short-term memory), nearest
neighbor, and decision trees all fall under this category.  On the broader scale of artificial intelligence, rule-based systems are also
mentioned.

Closely related to artificial intelligence are numerical analysis
methods.  While traditional methods such as interpolation and
approximation are used, heavy emphasis is placed on optimization
techniques such as Gauss-Newton or Kalman filters.
These are primarily used to diagnose components, which might be in part
due to the size of the components making such techniques feasible or due
to excessive size of machine learning models.

Another application area for machine learning is signal processing,
to which we counted face and object detection, image processing,
``signal processing'' itself, as well as particle filters.
Particularly for face and object detection and more general image
recognition, neural networks such as YOLO are used, which we counted
towards both.
Unlike numerical analysis, there is no clearly pronounced peak for signal processing.

While advertised as a main feature of digital twins if going by
\acite{grieves2017digital}, simulation and visualization appear
significantly less often as means for diagnosis.
One potential explanation would be that visualizations ``only'' serve
the purpose of informing humans, whereas machines could more easily work
with data in non-image form.  Another reason could be that
simulations are often intermediate steps -- for fault detection, one
would have to then interpret the simulation results by applying
human or artificial intelligence.

Other methods are only minor in occurrence and thus more susceptible
to noise than proper analysis.

Having established the models we wish to create, the diagnosis steps
we want to perform, and the methods to achieve them,
we still need to store all relevant data.
Sadly, information regarding this storage is often lacking:
storage remains unknown for \storage[unknown] out of
\total[records] \records{} even with liberal interpretations.
When data storage is mentioned or alluded to, nebulous entities such as the cloud (\storage[cloud]), fog (\storage[fog]) and
network edge (\storage[edge]) have precedence over
on-device (\storage[device]) or database (\storage[database])
storage.
Note that a single digital twin might be spread across multiple
locations.  This spread does not just refer to how cloud
servers spread storage across the globe but also to the copying of
data between the network edge and cloud when regarding them as singular
entities.  Of the \storageN[edge] edge-based digital twins,
\storageN[edge and cloud] have their data backed up in the cloud.
Blockchain-based digital twins (\storage[blockchain]) are at the time
of writing still a note in some margin, but given the hype blockchain
experiences as the technology itself, one can expect more in the
future.

\section{Trends and Issues}\label{sec:trends}

While conducting this systematic mapping study, we noticed three
major trends concerning digital twins for diagnosis in transportation.

Firstly, there is a strong relationship between digital twins and
artificial intelligence, in particular machine learning and its building
blocks, as well as some other technologies.
The reliance of digital twins on artificial intelligence (more narrowly
machine learning) has also been observed by
others~\cite{wu2020battery,wang2021application,qian2022digital},
along with data science~\cite{wu2020battery} or
big data~\cite{wang2021application,qian2022digital},
cloud computing~\cite{wang2021application} and the
Internet of things~\cite{wu2020battery}.
With the applications of digital twins currently mostly being limited to
monitoring it is difficult to say whether these technologies serve the
purpose of creating a digital twin or whether the digital twin serves
the justification of their use.
A more in-depth discussion of the role of artificial intelligence,
machine learning and big data are given in \cite{rathore2021role}.

Secondly, a fair number of digital twins are created in hope that they
help improve the current climate crisis.
A rough sketch of how this could be done is given by
\acite{grieves2017digital}: instead of
wasting resources in fruitless attempts at completing some task, with
adequate information, we can perform the task wasting fewer resources.
While this assumption is likely to hold in terms of financial resources,
more work needs to be done to demonstrate it in environmental terms.
In particular, as computational power, storage, and bandwidth increase,
so too are energy consumption, and related emissions increased and
therefore need to be taken into account.
This particularly concerns the intersection of digital twins with other
supposedly environment-friendly technologies such as electric cars.
While indeed the electrification of cars is necessary to curb emissions
and digital twins may well aid in the reuse of their batteries, we
can not hope for an ever-increasing fleet of electric and autonomous
cars to be a good strategy towards long-term sustainability.

Thirdly, digital twins raise some issues regarding privacy and security,
some of which have already been pointed out in prior
work~\cite{errandonea2020digital,bhatti2021towards,
  gao2021digital,qian2022digital}.
Few papers discuss these issues, and those that do contain a wide variety
of statements from claims that
edge computing or blockchain provide necessary levels of security or
privacy~\cite{broekman2021real,kumar2018novel,marai2021roads} over other
potential solutions within their specific application
\cite{campolo2020digital} to
delegations towards future work~\cite{anandavel2021application,
  tihanyi2021towards,zhao2020intelligent} and
mere acknowledgments~\cite{clemen2021multi,shim2019digital,
  wang2022mobility}.

Overall, the state of information security among digital twins is not a
satisfactory one.  Even the offered solutions shy away from
mentioning concerns they don't address.
For example, \cite{broekman2021real} claims that not storing media
footage addresses privacy concerns, but malicious software or hardware
within the very same devices could store or transmit said data and
benign software still transmits metadata that might find their way
into exploits.  Similarly, \cite{campolo2020digital} uses ``properly
configured views'' for ``authorized applications'', but at the same
time permits the cloud platform to derive whatever analytics it deems
fit.

\acite{wang2022mobility} contextualize cybersecurity and privacy risks
by opposing public and private cloud platforms.  They define a private
cloud platform as one consisting of resources used exclusively by one
business or organization and a public cloud platform as one that
lacks this exclusivity.
This paints a misleading picture in which private cloud platforms only
operate on data internal to that organization (and thus personal),
whereas public platforms can mix and match data from many organizations, thus making it as good as public.
In practice, it is first of all questionable whether that data is,
in fact, internal to the organization or taken from people who are not
offered any meaningful choice of consent.
Further, these cloud platforms serve primarily the (mostly financial)
interests of those willing to host them.
As of the time of writing, none of the platforms that have the
capabilities of hosting a digital twin are under public control or
serve the public interest, let alone both.

\section{Conclusion}\label{sec:conclusion}

We conducted a systematic mapping study of digital twins for
diagnosis in transportation.
We found that (at least) \total[papers] papers have been published
discussing digital twins as a means of performing diagnosis in
transportation systems (\ref{rq1}).

We mapped these digital twins to different types (\ref{rq2})
according to their underlying models, diagnostic use and diagnosis
methods.
Most papers suggested using artificial intelligence for diagnosis,
which typically meant artificial intelligence based on machine learning.
As for diagnostic use, most digital twins served only monitoring or
fault detection, with little research on localization or
mitigation.  Future work could aid in closing this gap.

Besides trends related to artificial intelligence, we noticed the
climate impact of transport systems as an issue digital twins aim to
aid in solving, as well as new challenges regarding
privacy and security (\ref{rq3}).

\section*{Acknowledgment}
This paper is part of the AI4CSM project that has received funding within the ECSEL JU in collaboration with the European Union's H2020 Framework Programme (H2020/2014-2020) and National Authorities, under grant agreement No.101007326. The work was partially funded by the Austrian Federal Ministry of Climate Action, Environment, Energy, Mobility, Innovation and Technology (BMK) under the program ``ICT of the Future'' project 877587.

\phantomsection 
\printbibliography

\end{document}